\documentclass{iaus}
\usepackage{graphicx}

\title[Secular Resonances in the Planetary Systems] 
{On  Secular Resonances of Small Bodies in the Planetary Systems}

\author[Ji et al.]   
{Jianghui JI$^{1,2}$%
  \thanks{Present address: Department of Terrestrial Magnetism, Carnegie
Institute of Washington, 5241 Broad Branch Road NW, Washington, DC
20015-1305},
 L. LIU$^3$  \and  G.Y. LI$^{1,2}$}

\affiliation{$^1$Purple  Mountain  Observatory, Chinese  Academy
of  Sciences,  Nanjing  210008, China \break email: jijh@pmo.ac.cn\\[\affilskip]
$^2$National Astronomical Observatory, Chinese Academy of
Sciences, Beijing 100012, China \break email: xhliao@nju.edu.cn\\[\affilskip]
$^3$Department of Astronomy,  Nanjing University, Nanjing  210093,
China}

\pubyear{2006}
\volume{236}  
\pagerange{1--8}
\jname{Near Earth Objects, Our Celestial Neighbors: Opportunity
and Risk}
\editors{A. Milani, G.B. Valsecchi, \& D. Vokrouhlicky, eds.}
\begin{document}

\maketitle

\begin{abstract}
We investigate the secular resonances for massless small bodies
and Earth-like planets in several planetary systems. We further
compare the results with those of Solar System. For example, in
the GJ 876 planetary system, we show that the secular resonances
$\nu_1$ and $\nu_2$ (respectively, resulting from the inner and
outer giant planets) can excite the eccentricities of the
Earth-like planets with orbits 0.21 AU $\leq  a  <$ 0.50 AU and
eject them out of the system in a short timescale. However, in a
dynamical sense, the potential zones for the existence of
Earth-like planets are in the area 0.50 AU $\leq a \leq$ 1.00 AU,
and there exist all stable orbits last up to $10^5$ yr with low
eccentricities. For other systems, e.g., 47 UMa, we also show that
the Habitable Zones for Earth-like planets are related to both
secular resonances and mean motion resonances in the systems.

\end{abstract}

\firstsection 

\section{Introduction}
Since the discovery of the first Jupiter-mass planet orbiting the
solar-type star 51 Peg (Mayor \& Queloz 1995), it has been more
than a decade. The breakthrough of scientific finding not only
arouses great interests to search for other habitable planets or
alien civilization worlds outside our own solar system, but also
explicitly confirms that the planets can be at birth anywhere
about their parent stars in the circumstellar disks, because these
flat disks enshrouding young stars are considered to be a common
feature of stellar evolution and of planetary system formation
(Beichman et al. 2006). Primordial protoplanetary disks contain
gas and dust and supply the raw ingredients from which the new
planetary systems can form. To date, over 160 planetary systems
(see also http://www.exoplanets.org and http://exoplanet.eu/) have
been discovered by the measurements of Doppler radial velocity and
other observational methods (Butler et al. 2006). More than 200
extrasolar planets have been detected about solar-type stars.
Currently, amongst the detected systems, there are 20
multiple-planet systems, e.g., two-planet systems (e.g.,47 UMa
etc.), three-planet systems (e.g., Upsilon And etc.) and a
four-planet system, 55 Cancri. Then, the studies of the dynamics
or formation of these systems are essential to understand how two
(or more) planets originate from and evolve therein. In recent
years, many authors have investigated the dynamical evolution of a
planetary system and intended to reveal the possible mechanisms
that stabilize a system, especially for those involved in the mean
motion resonances (MMR), e.g., 2:1 MMR (GJ 876, HD 82943, HD
128311, HD 73651), and explored the secular interactions in the
multiple systems (e.g., Hadjidemetriou 2002, 2006; Gozdziewski
2002, 2003; Ji et al. 2002, 2003; Lee \& Peale 2002, 2003).
Herein, we investigate the secular resonances for massless small
bodies and Earth-like planets in several planetary systems, which
is extremely important to make clear what dynamical structure of
the newly-discovered systems may hold and how the secular
resonances would have influence on the motions of the potential
Earth-like planets and the location of the Habitable Zones (HZ).
The HZs are generally believed as suitable locations where the
biological evolution of life is able to develop on planetary
surfaces in environment of liquid-water, subtle temperature and
atmosphere components of CO$_{2}$, H$_{2}$O and N$_{2}$ (Kasting
et al. 1993). In the meantime, the planetary habitability is also
relevant to the stellar luminosity and the age of the star-planet
system (\cite{Cun03}). However in Solar System, it is believed
that the asteroids in the main belt can undergo secular resonances
with respect to Jupiter (or Saturn), and their eccentricities can
be greatly excited. The bodies can cross and approach Earth in
million years, as a near-Earth object (NEO). Hence, the present
study is mainly focusing on the issue that such secular resonances
make a difference in other planetary systems.

Moreover, the scientific objectives of several space missions
(e.g., SIM, TPF\footnote{http://planetquest.jpl.nasa.gov/SIM, and
http://planetquest.jpl.nasa.gov/TPF}) will be in part contributed
to be hunting for the Earth-like planets, although this may come
true after a significant improvement of precision of ground
observations. Then, we also start such studies in the planetary
systems advancing these projects. At first, we will quickly review
the secular resonances taking place for the asteroids in the main
belt.

\section{Secular Resonances for Main Belt Asteroids in Solar System}

It is well-known that the concentration or depletion for the
asteroids in the main belt are associated with the mean motion
resonances (MMR) with Jupiter's orbit and secular resonances
(Williams 1969; Milani \& Knezevic 1992; Morbidelli \& Moons
1993). The main belt asteroids are populated at the 3:2, 4:3 and
1:1 MMR with Jupiter, but rarely resided in the 2:1, 3:1, 5:2 and
7:3 resonant regions, which are called the Kirkwood gaps.
Moreover, the secular resonances are responsible for the long-term
dynamical evolution for small bodies. In general, secular
resonances occur when the longitude of the perihelion or that of
the ascending node of the small body shares the same precession
rate as that of the massive giant planet (e.g. Jupiter and
Saturn). There are three governing secular resonances in the
asteroidal belt, known as the $\nu _5$, $\nu _{6}$ and $\nu _{16}$
resonances. The formed two are called apsidal secular resonances
with respect to Jupiter and Saturn, respectively and can pump up
the eccentricity of a small object; the latter one is the nodal
secular resonance with respect to Saturn, which can enhance the
inclination of the body. At present, it is believed that the NEOs
are principally considered to be objects ejected from the main
belt through a complicated dynamical process, where mean motion
resonances as well as secular resonances play a vital role in
their dynamical transportation (Morbidelli \& Moons 1993;
Froeschl{\' e} 1997; Morbidelli et al. 2002), indicating that the
overlapping of mean motion resonances and secular resonances
(Morbidelli \& Moons 1993; Moons \& Morbidelli 1995) can lead to
large chaotic zones for the relevant asteroids.

For example, the small bodies trapped in a 3:1 orbital resonance
with Jupiter (occupying the semi-major axes $\sim$ 2.5 AU) are
rarely distributed, involved in Kirkwood gaps. Wisdom (1983)
pointed out that the chaotic motion for the asteroids in 3:1 MMR
can increase the eccentricities and then make them approach and
intersect the orbit of Mars (even Earth).  Herein, Figure 1 shows
that the orbital evolution for a massless test particle, however,
over the time span of (0.65 Myr, 0.80 Myr) due to $\nu _{5}$ and
3:1 resonance, the eccentricity $e$ of the small body is excited
up to 0.60 and meanwhile the semi-major axis $a$ drops down to
2.20 AU, being an Earth-crossing body.  In other numerical
investigations for the dynamical evolution of the minor bodies
over millions of years, we also find that several NEOs can be
temporarily locked a 3:1 orbital resonance and also experience
secular resonance $\nu_{5}$ (or $\nu_{6}$) with Jupiter (or
Saturn), then confirm that the 3:1 orbital resonance and secular
resonances play an important role in the origin for NEOs by
previous studies (e.g., Morbidelli et al. 2002 and references
therein).

\begin{figure}
\includegraphics[height=2in, width=3in]{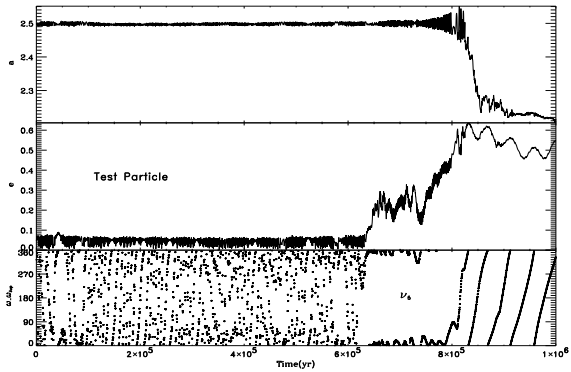}
  \caption{The time behavior of the semi-major axis $a$, the
eccentricity $e$ and $\varpi - \varpi_J$ for the test particle.
$a$ slightly oscillates about 2.50 AU within 0.6 Myr, over the
time span of (0.65 Myr, 0.80 Myr) due to $\nu_5$ resonance (see
bottom and middle panels), and $e$ is excited above 0.60, while
$a$ goes down to 2.20 AU. Eventually, the test body becomes a NEO
candidate.} \label{fig1}
\end{figure}

\begin{table}
\def~{\hphantom{0}}
  \begin{center}
   \caption{Properties of 2 multiple planet systems (data adopted from
    Laughlin \& Chambers 2001; Fischer et al. 2003)
    }
  \label{table1}
\begin{tabular}{lccccc}
\hline

Planet  &$M_{star}$($M_{\odot}$)&$M$sin$i$($M_{Jup}$) &Period
$P$(days) & $a$(AU)  &ecc.\\

\hline
GJ  876  b  &0.32 &3.39  &62.09  &0.211  &0.05 \\
GJ  876  c  &0.32 &1.06  &30.00  &0.129  &0.31 \\
47  UMa  b  &1.03 &2.86  &1079.2 &2.077  &0.05 \\
47  UMa  c  &1.03 &1.09  &2845.0 &3.968  &0.00 \\
\hline
\end{tabular}\normalsize
\end{center}
\end{table}

\section{Secular Resonances in  Extrasolar  Systems}
In order to investigate the dynamical structure or Habitable Zones
in the planetary systems, we also performed extensive numerical
simulations for the planetary configurations of two giant planets
with one fictitious low-mass body  for several systems (e.g., 47
UMa, GJ 876, etc). We also show that the secular resonances can
affect the motions of the small bodies in these systems, and shape
the dynamical architecture in the debris disk as mean motion
resonances. As for the methodology, we use a N-body code (Ji et
al. 2002) of direct numerical simulations with the RKF7(8)
(Fehlberg 1968) and symplectic integrators (Wisdom \& Holman
1991). We always take the stellar mass and the minimum planetary
masses from Table 1, while the mass of an assumed terrestrial
planet is adopted to be in the range from 0.1 $M_{\oplus}$ to 10
$M_{\oplus}$. The used time stepsize is usually $\sim$ 1\%-2.5\%
of the orbital period of the innermost planet. In addition, the
numerical errors were effectively controlled over the integration
timescale. The typical integration timescale for the simulation is
$10^5$ yr. The main results now follow.
\subsection{GJ 876}
\begin{figure}
\includegraphics[height=2in, width=2.9in]{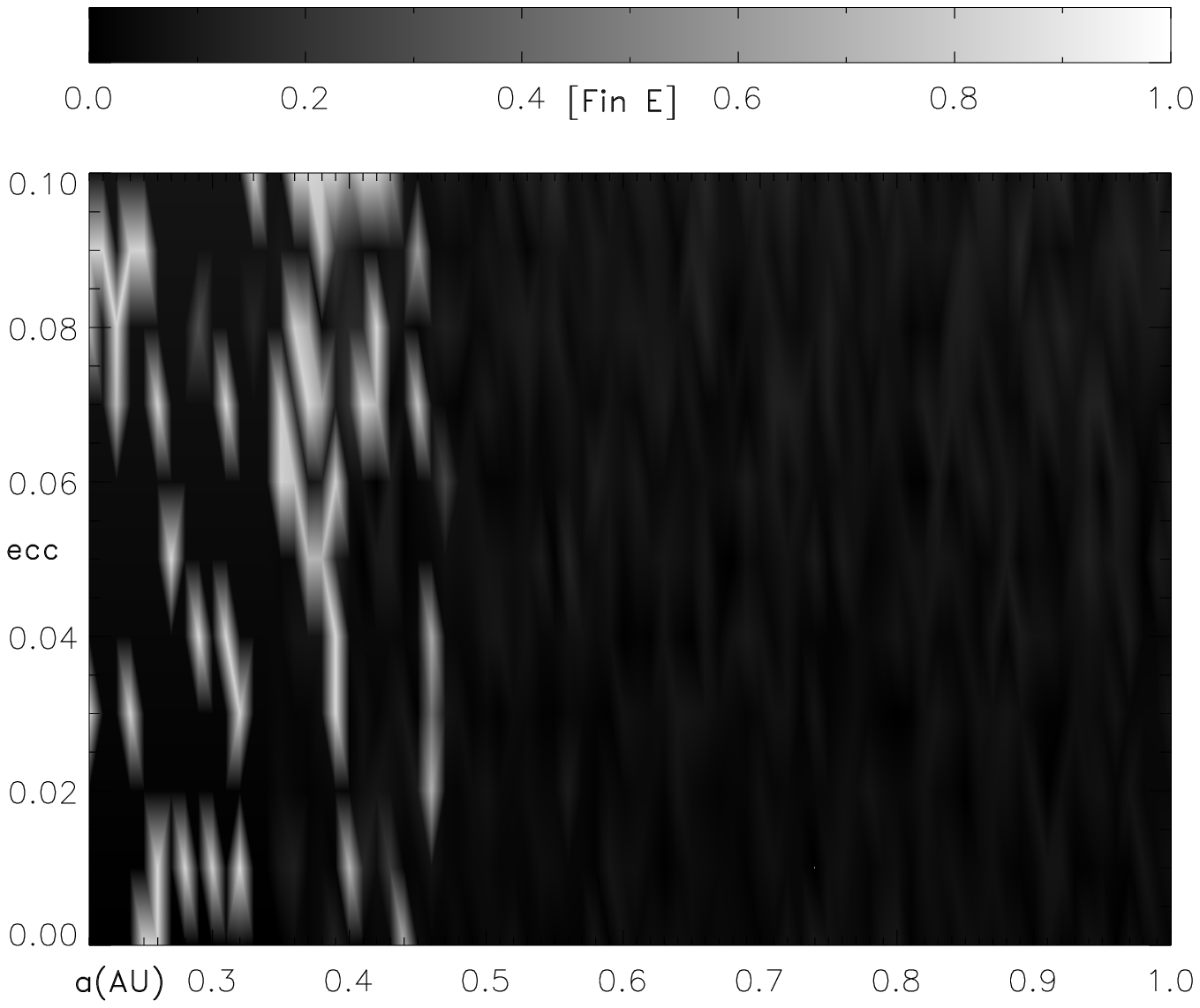}
\includegraphics[height=2in, width=2.9in]{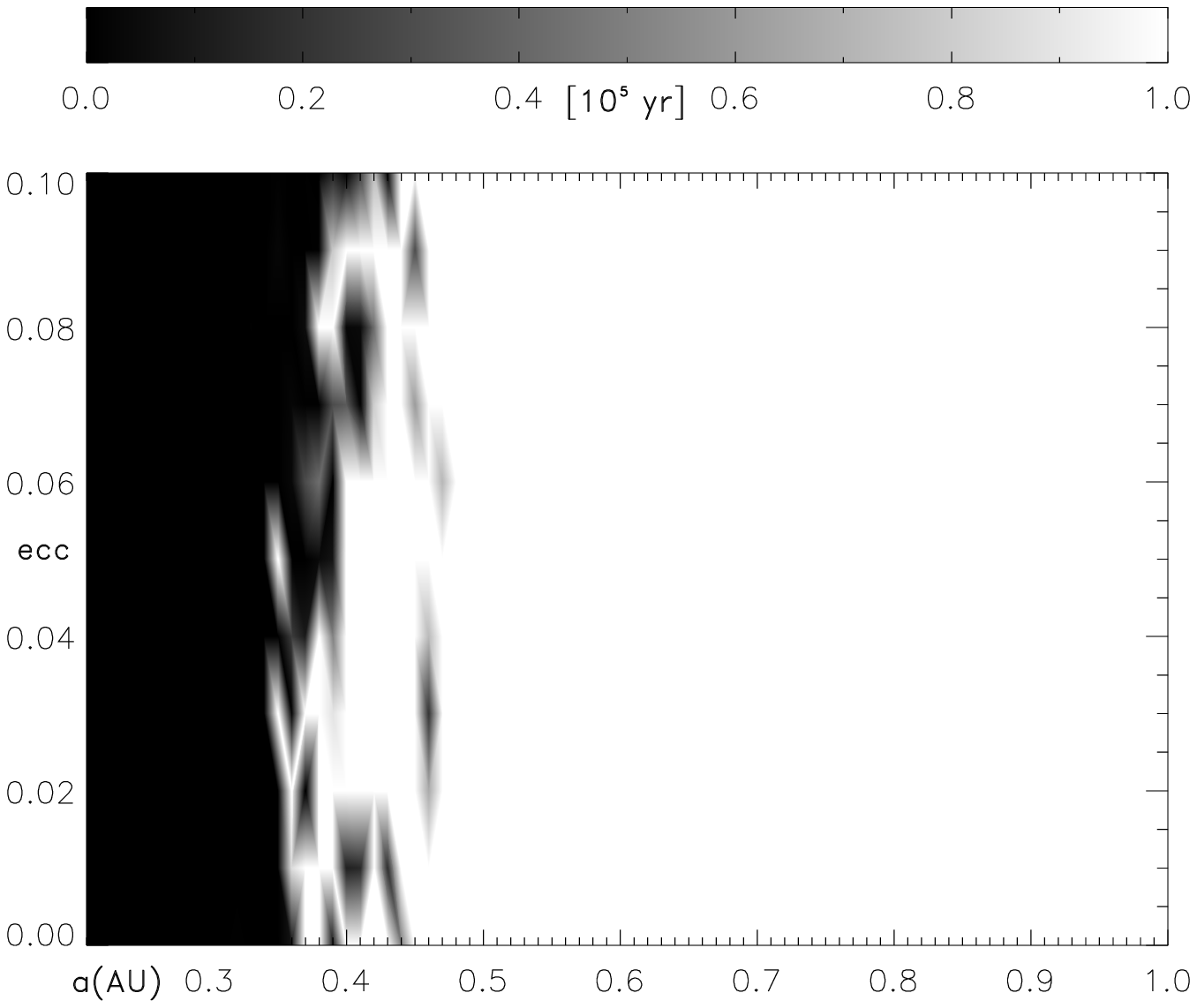}
  \caption{\textit{Left panel}: Contour of the final eccentricities for the Earth-like planets
   in GJ 876 system.
   Horizontal and vertical axes are the initial values of $a$ and $e$.
   In the region 0.21 AU $\leq a < $ 0.50 AU, the eccentricities can
   be pumped up to high values $\sim$ 1 or these bodies are
   directly ejected from the system due to the starting dynamical instability.
   Hence, in this region, the Earth-like planets are strongly chaotic and
   cannot survive in the system.
    \textit{Right panel}: Surviving time for Earth-like planets in the system.
    The Earth-like planets evolve with short dynamical time before they
    end their destinies in the area 0.21 AU  $\leq  a < $ 0.30 AU,
    indicating that these orbits are completely unstable, for the initial conditions.
    The chaotic behaviors of the Earth-like planets in 0.21 AU $\leq  a < $ 0.50 AU  are,
    somewhat related to  two secular resonances ($\nu_{1}$ and
    $\nu_{2}$).
} \label{fig2}
\end{figure}

The M dwarf main-sequence star GJ 876 with an estimated mass of
0.32 $M_{\odot}$ is the lowest mass star that hosts planets, and
two Jupiter-like planets (Marcy et al. 2001) are revealed with
minimum masses of 1.89 $M_{Jup}$ and 0.56 $M_{Jup}$ in this
system. Moreover, the ratio of the orbital periods of two planets
is close to a mean motion commensuration of 2:1. Being the first
discovered 2:1 resonant planetary system, the GJ 876 has generated
great interests and the long-term dynamics and planetary formation
for two giant companions are extensively investigated (e.g.,
Hadjidemetriou 2002; Ji et al. 2002; Lee \& Peale 2002; Beauge \&
Michtchenko 2003; Kley et al 2005; Laughlin et al. 2005, and
references therein). However, the planetary formation theory
(Lissauer 1993) suggests that even low-mass planets (e.g.,
Earth-like planets) may exist about the most abundant M dwarf
stars with mass of 0.08 - 0.8 $M_{\odot}$, which covers 75\% of
the total stellar population in the galaxy. For instance, Butler
et al. (2004) announced the discovery of a Neptune-mass planet
($\sim 18$ $M_{\oplus}$) about M dwarf star GJ 436, implying the
potential existence of the terrestrial or Neptunian planet in
other systems.

Thus, we exhaustively investigated the case of two
coplanar-configuration giant companions with one terrestrial
planet. The initial orbital parameters were adopted as follows:
the low-mass terrestrial bodies were placed at an equal interval
of 0.01 AU from 0.21 AU to 1.0 AU in $a$, the eccentricities $e$
were taken every 0.01 from 0 to 0.1, the inclinations are
$0^{\circ} < I < 5^{\circ}$, and the other angles were randomly
distributed between $0^{\circ}$ and $360^{\circ}$. Then each
integration was carried out for $10^{5}$ yr.

The numerical outcomes reveal that the two secular resonances
$\nu_{1}$ and $\nu_{2}$ respectively arising from the inner and
outer giant planets are responsible for the chaotic motions of the
Earth-like planets. To understand the vital role of the secular
resonances, we have carried out several computations. If a
terrestrial planet has the mass of 1 $M_{\oplus}$, the location
for $\nu_{2}$ secular resonance is $\sim 0.4550$ AU, where the two
eigenfrequencies for the terrestrial body and the outer giant
planet are provided by Laplace-Lagrange secular theory (Murray \&
Dermott 1999) are, respectively,  $1^{\circ}.83$ yr$^{-1}$  and
$1^{\circ}.90$ yr$^{-1}$. This is fairly in agreement with
numerical results, where Figure 2 (\textit{left panel}) shows the
excitation of eccentricity of the Earth-like planets at $\sim
0.45$ AU. In addition, the relevant location for $\nu_{1}$ secular
resonance is $\sim 0.2930$ AU, in this case the terrestrial planet
almost shares the same eigenfrequency as the inner giant planet,
the values are $21^{\circ}.30$ yr$^{-1}$ and $20^{\circ}.94$
yr$^{-1}$, respectively.  The mutual Hill radius is
$R_{H}={[(M_{1} + M_{2})/(3M_{s})]}^{1/3}(a_{1} + a_{2})/2$, where
$M_{s}$, $M_{i}$ are the masses of the host star and the planets
(the subscript $i=1, 2$, stands for the inner and outer planets,
respectively), and $a_{1}$, $a_{2}$ the semi-major axes. Using the
data in Table 1, we obtain $3R_{H}=0.084$ AU, where $a_{2}=0.211$
AU, then we have an exterior influence boundary $\sim0.295$ AU,
which is almost equal to $\nu_{1}$. Thus, the Earth-like planets
with orbits $\sim 0.30$ AU are strongly affected by $\nu_{1}$, and
this also confirms our numerical explorations. On the other hand,
plenty of mean motion resonances exist and overlap within Hill
radius. The dynamical lifetime of the bodies will decrease
drastically.

In a dynamical sense, for GJ 876, the potential existence of the
Earth-like planets\footnote{Rivera et al. (2005) reported a $\sim
7.5$ Earth-Mass planet about GJ 876, with the orbital period of
1.938 day, and they also indicated that additional planets may be
revealed in this system with more observational data.} concerns
the region 0.50 AU $\leq a \leq 1.00$ AU. Stable orbits exist, up
to $10^5$ yr with low eccentricities (see \textit{right panel} of
Figure 2) in the resulting evolution. This is because the initial
orbits of Earth-like planets are a bit far away from the two
secular resonances, free from secular perturbation. Moreover, the
dynamical stability beyond 0.50 AU also suggests that outer belts
for unaccreted planetesimals may exist in this system.

The formation of two giant planets in the GJ 876 system has been
recently modelled by Lee \& Peale (2002) and Kley et al. (2005),
and the planets were likely captured into the 2:1 resonance by
converging differential migrations in the protoplanetary disk. In
this sense, the motions of the Earth-like planets or the
planetesimals in the disk may be influenced by the orbital
migration of the two giant planets, and they may be swept out
directly or captured into the resonance with the two larger
planets of GJ 876. This should be re-examined in future studies.

\subsection{47 UMa}
The main sequence star 47 UMa is of spectral type G0 V with a mass
of 1.03$M_{\odot}$. Butler \& Marcy (1996) reported the discovery
of the first planet in the 47 UMa system which has become one of
the most amazing systems particularly after the subsequent release
of an additional companion (Fischer et al. 2002, 2003). It is
sometimes thought to be a close analog of our own solar system:
for example, the mass ratio of the two giant companions in 47 UMa
is $\sim$ 2.62 (Table 1), as comparable to that of Jupiter-Saturn
(JS) of 3.34; and the ratios of the two orbital periods are very
similar. Hence, one may wonder whether there exists additional
members in 47 UMa system (see Ji et al. 2005 for details).
\begin{figure}
\includegraphics[height=2in, width=2.6in]{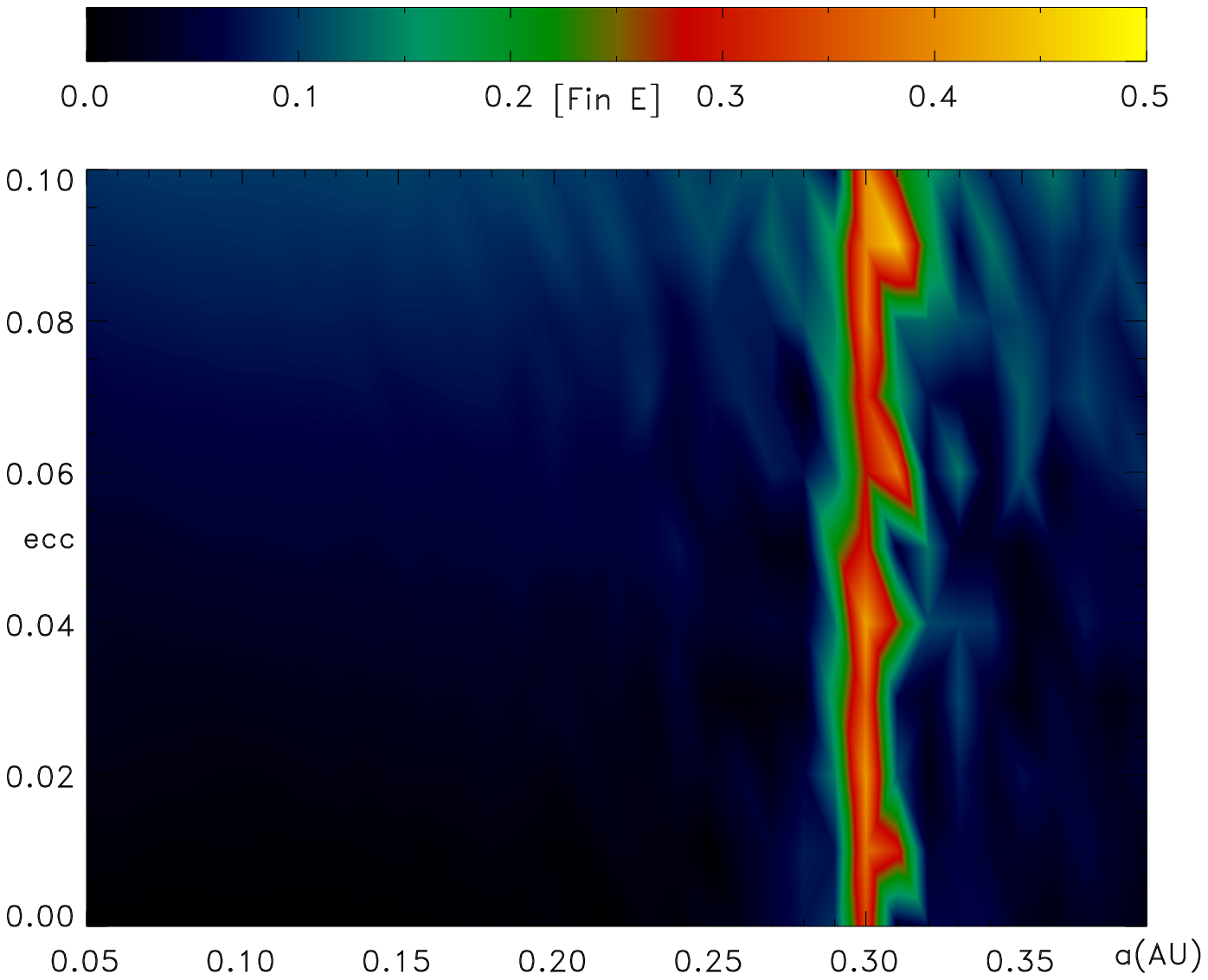}
\includegraphics[height=2in, width=2.6in]{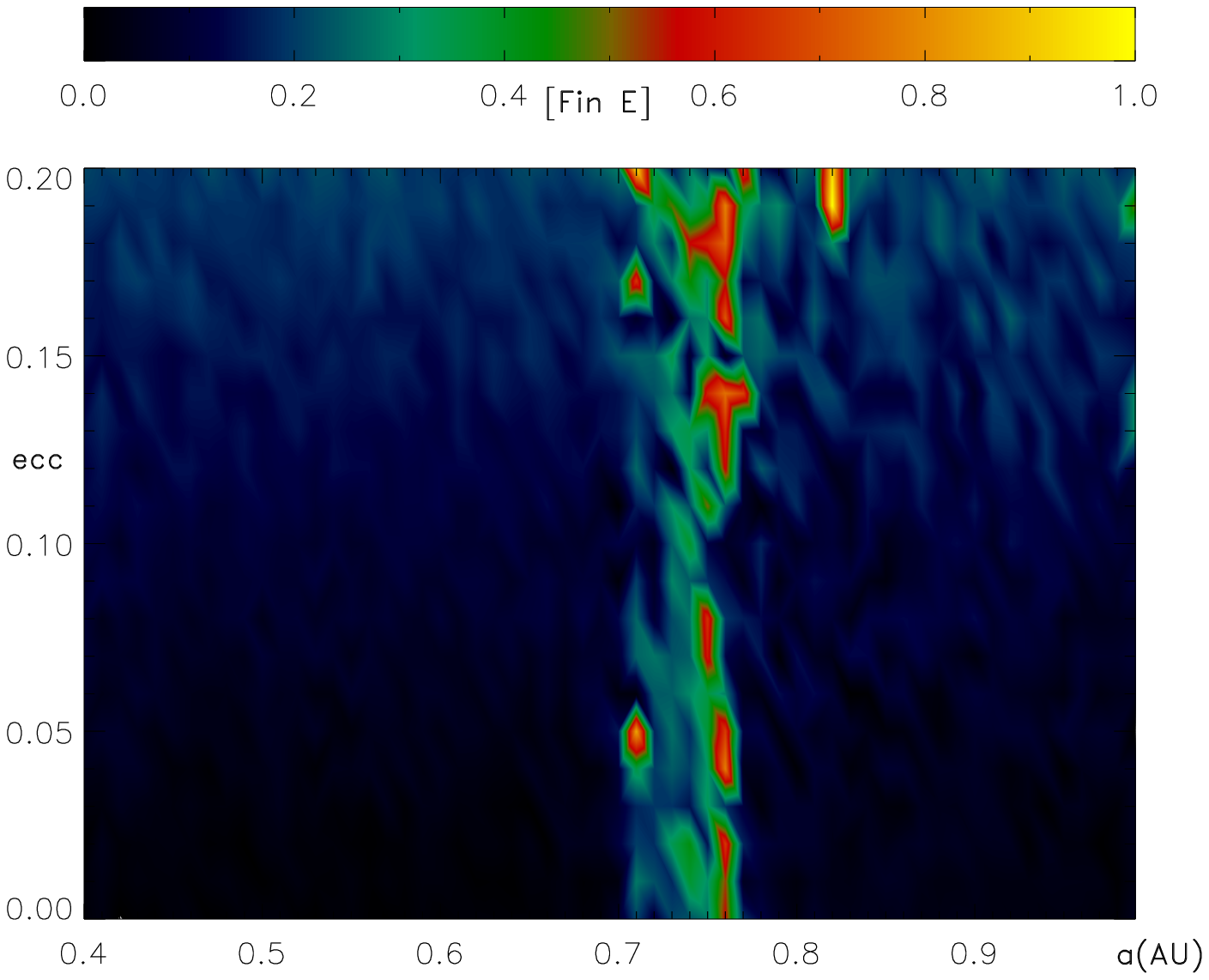}
  \caption{The contour of status of the final eccentricities
  for Earth-like planets, the vertical axis for the initial $e$.
  \textit{Left panel}: 0.05 AU  $\leq a <$ 0.4 AU for 1 Myr.
  Notice that the secular resonance at $\sim$ 0.30 AU pumps up
  the eccentricities.
  \textit{Right panel}: 0.4 AU $\leq a <$ 1.0  AU for 5 Myr.
  The eccentricity $e$ of the orbits with 0.70 AU $< a <$ 0.78  AU
  is excited and in the 2:9 MMR  at $\sim$ 0.76 AU, $e$ can reach
  $\sim$ 0.90.
} \label{fig3}
\end{figure}

\cite{Lau02} and Gozdziewski (2002) studied the long-term
stability of 47 UMa and pointed out that the secular apsidal
resonance can help stabilize the two giant planets in an aligned
configuration with the libration of their relative periapse
longitudes (Ji et al. 2003). Then the eccentricities avoid larger
oscillations due to this mechanism, as a result, this system can
even survive for billion years (Barnes \& Quinn 2004). Several
pioneer works were concentrated on the structure of the system and
presented a preliminary understanding of this issue. Jones, Sleep
\& Chambers (2001) investigated the existence of Earth-mass
planets in the presence of one known giant planet, and
subsequently \cite{Lau02} and Asghari et al. (2004) further
studied the stability of massless test particles about the
so-called Habitable Zones (HZ) according to some earlier solutions
(Fischer et al. 2002), where the dynamical model was treated as a
restricted multi-body problem. Nevertheless, as the terrestrial
planets possess significant masses, they can interact with the two
giant planets by mutual gravitation, which may result in secular
effects for the planetary system. Accordingly, we should take into
account the masses of terrestrial bodies in the model when
exploring the dynamical architecture. Herein, we performed
extensive simulations to examine the dynamical architecture in
both the HZ and the extended areas, for Earth-like planets (with
masses from 0.1 $M_{\oplus}$ to 10 $M_{\oplus}$) of 47 UMa with
stable coplanar planetary configuration, based on the best-fit
orbital parameters given by Fischer et al. (2003). These new
reliable orbital solutions are derived from additional follow-up
observations, hence they can represent the actual motions of the
system under study. On the other hand, the discovery of the
close-in Neptune-mass planets (Butler et al. 2004; McArthur et al.
2004; Santos et al. 2004) demonstrates that it may be possible for
less massive planets ($\sim M_{\oplus}$) to move close to the
star. Therefore, we also explored low-mass planets in the region
0.05 AU $\leq a < 0.4$ AU and we found that the secular resonance
arising from the inner giant planet render the eccentricity
excitations for the Earth-like planets.

For 0.05 AU  $\leq a <$ 0.4 AU,  we explored the secular evolution
of hundreds of "hot Earths" or "hot Neptunes" over timescale of 1
Myr. All the simulations are dynamically stable for $10^{6}$ yr,
and 96\% of the orbits posses $e_{final}< 0.20$. However, Figure 3
(\textit{left panel}) shows that the eccentricities for the bodies
at $\sim 0.30$ AU are excited to $\sim 0.40$, where the secular
resonance $\nu_{1}$ ($41^{"}.11$ yr$^{-1}$) of the inner giant
planet (similar to $\nu_{5}$ for Jupiter) is responsible for
excitation of the eccentricities. In addition, Malhotra (2004)
also presented similar results, showing that the eccentricities of
massless bodies are excited in the debris disk at $\sim 0.30$ AU,
using nonlinear analytical theory.

In the region 0.4 AU $\leq a <$ 1.0  AU, more than one thousand of
simulations were carried out for 5 Myr each (see \textit{right
panel} in Figure 3). Most of the Earth-like planets about 1:4 MMR
at $\sim 0.82$ AU move stably in bounded motions with
low-eccentricity trajectories, except for two cases where the
eccentricities eventually grow to high values. The secular
resonance $\nu_{2}$ arising from the outer companion (similar to
$\nu_{6}$ for Saturn) can remove the test bodies. Herein $\nu_{2}$
can also influence the Earth-like planets in this system. The
terrestrial planets that all bear finite masses that may change
the strength of this resonance; on the other hand, the location of
the secular resonance is changed due to the orbital variation of
the outer companion. For a terrestrial planet with a mass of $10
M_{\oplus}$, the location for $\nu_{2}$ secular resonance is about
$\sim 0.70$ AU, where the two eigenfrequencies for the terrestrial
body and outer giant planet given by the Laplace-Lagrange secular
theory are, respectively, $211^{"}.37$ yr$^{-1}$ and $225^{"}.48$
yr$^{-1}$. This indicates that both planets have almost the same
secular apsidal precession rates in their motion. Hence, the
$\nu_{2}$ resonance, together with the mean motion resonance, can
work at clearing up the planetesimals in the disk (see Fig. 3) by
the excitation of the eccentricity (see also \cite{Nag05}).

We point out that the most likely candidate for habitable
environment is  terrestrial planets with orbits in the ranges 0.8
AU $\leq a < 1.0$ AU  with low eccentricities (e.g., $0.0 \leq e
\leq 0.1$). However, in our own solar system there are no
terrestrial planets from the 1:4 MMR out to Jupiter, although
there are stable orbits there. This may suggest that although some
orbits are stable, the conditions are such that terrestrial
planets cannot form so close to giant planets. Perhaps this is
because runaway growth is suppressed due to the increased
eccentricities from the perturbations of the giant planet. In 47
UMa, the corresponding region runs from 0.82 AU on out (see Fig.
3), almost completely covering the HZ. Hence, it would be
reasonable to conclude that the only proper place to find
habitable planets in this system would be at about 0.80 AU. But
this should be carefully examined by forthcoming space
measurements (e.g., SIM or GAIA) capable of detecting low-mass
planets.

\section{Summary and Discussion}
In this work, we investigate the secular resonances for massless
small bodies and Earth-like planets in several planetary systems
with two giant planets (e.g., 47 UMa and GJ 876) by extensive
numerical simulations, and further we have studied the potential
existence of Earth-like planets in the related regions for these
systems. In final, we summarize the following results:

(1) We can see that the 47 UMa planetary system may be a close
analog of our solar system, and even it can also own several
terrestrial members resembling the inner solar system (Ji et al.
2005). Besides, the two giant planets in the 47 UMa are similar to
the Jupiter-Saturn pair in the solar system, and the corresponding
secular resonances originating from them can stir the low-mass
small bodies with low eccentricities in the initial "cold" disk.
As to other systems, we also find that the Habitable Zones for
Earth-like planets are related to both secular resonances and mean
motion resonances in these systems, which may play an important
role of shaping the asteroidal belts. A comparative study has been
also performed in other planetary systems (see {\' E}rdi et al.
2004; Dvorak et al. 2004; Ji et al. 2006) with one or more giant
planets to explore whether Earth-like planets can exist there, and
further to locate less massive undetected planets or characterize
the nature of the potential asteroidal structure in general
planetary systems.

(2) The habitability for the development of biological evolution
depends on many factors, such as the liquid water state, the
temperature constraint, the atmosphere composition, the obliquity
and rotation rate of a target terrestrial planet, the stellar
luminosity, etc. However, in a dynamical viewpoint, it also
requires that the habitable terrestrial planets have stable orbits
in the HZ with low eccentricity at a nearly circular trajectory,
herein we show that the secular resonances can excite some orbits
residing in the HZ of a system, which may provide useful
information or place some constraints on the observational
strategy to discover such low-mass planets.

\begin{acknowledgments}
We are grateful to the referee, Anne Lemaitre for the valuable
suggestions that help to improve the contents. This work is
financially supported by the National Natural Science Foundations
of China (Grants 10573040, 10673006, 10203005, 10233020) and the
Foundation of Minor Planets of Purple Mountain Observatory.
\end{acknowledgments}

\end{document}